# LECTOR Based Clock Gating for Low Power Multi-Stage Flip Flop Applications

Pritam Bhattacharjee, Bipasha Nath, Alak Majumder[†]
*VLSI Design Laboratory,*
*Department of Electronics & Computer Engineering,*
*National Institute of Technology, Arunachal Pradesh,*
*Yupia, Dist.- Papumpare, Arunachal Pradesh, India.*
[†]*majumder.alak@gmail.com*

**Abstract**

*Power dissipation in integrated circuits is one of the major concerns to the research community, at the verge when more number of transistors is integrated on a single chip. The substantial source of power dissipation in sequential elements of the integrated circuit is due to the fast switching of high frequency clock signals. These signals do not carry any information and are mainly intended to synchronize the operation of sequential components. This unnecessary switching of Clock, during the HOLD phase of either 'logic 1' or 'logic 0', may be eliminated using a technique, called Clock Gating. In this paper, we have incorporated a recent clock gating style called LECTOR–based clock gating (LG–CG) to drive multi–stage architecture and simulated its performance using 90nm CMOS Predictive Technology Model (PTM) with a power supply of 1.1V at 18GHz clock frequency. A substantial savings in terms of average power in comparison to its non–gated correspondent has been observed.*

## 1. Introduction

The trend of low power applicability with increasing operating frequency in integrated circuit technology has brought in the necessity to search for intelligent power reduction techniques. It is known that major constituents of power consumption in chip design are the static power due to current leakage through the active and inactive devices and the dynamic power due to switching of the active devices. The current leak from the power supply ($V_{DD}$) to the ground known to be the contention current during logic transition in CMOS is inevitable. To solve this issue, it is needed to stop the leak through the power line. It can be done using scale down of the supply voltage, which on the other side, initiates the drop down of the operation speed as the supply voltage approaches the threshold voltage ($V_T$) of the transistors used in the design. The $V_T$ can be reduced but again reduction in $V_T$ will result in exponential increase in sub-threshold leakage current. In 2004, a technique based on the approach of effective stacking of PMOS and NMOS transistors between the $V_{DD}$ and ground called LECTOR has been reported [1]. This technique has the potential of leakage reduction through the power line sacrificing a little on the delay performance. On the other hand, switching power which is mainly governed by the clock net in the design is a major concern for dynamic power dissipation. For solving the switching power issue, the clock signal is shut off for the portion of the design, where no operations are taking place for a certain amount of time. In this way, switching activity factor of the design is reduced leading to an enormous saving on the part of dynamic power. This technique is known as Clock Gating [2, 3]. Circuit designers started using Clock Gating in their designs as it is a type of optimization in the usage of clock signals. The principle of clock gating is shown in figure 1. The system clock is gated using a gating logic suppressing its unnecessary switching in the idle time. Controlling of the clock gating logic (CGL) is done by the control signal.

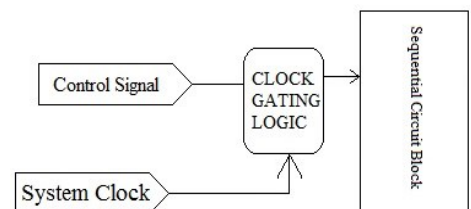

**Figure 1. Principle of Clock Gating.**

There are various architectures prevalently in clock gating with variety usage of the control signal. The most recent among them is the LECTOR–based clock gating reported in [4], where the pertaining test circuit used for justifying the gating logic was a simple D-Flip-flop. Though the sequential application incorporated by this gating logic has been reported in [12], but the question lies whether this gating logic is suitable for multi–stage operations, as it is important because in larger circuit topologies, the best trend is to cascade as many number of stages.

Therefore, in this paper, we have looked upon the performances of multi–stage circuits when introduced to the recent clock gating logic.

Organization of the paper is as follows: A brief survey on clock gating logics is explored in Section–2. In Section–3, the multi–stage circuit representation and its work ability is discussed. Then, in Section–4, we have observed the performance difference of gated and non–gated scenario of multi–stage test circuit and finally, in Section–5, we conclude our work.

## 2. Brief Survey on Clock Gating

The clock gating has various design styles. They are broadly categorized as Latch–free based gating, latch–based gating and Flip-Flop based gating and their pros and cons are extensively reported in [5–7]. There are also hybrid gating styles derived from the combination of latch–free based and latch–based gating. These hybrid styles are broadly categorized as Double–gated gating, NC²MOS gating, dynamic gating and bootstrap XOR gating as reported in [9], [10] and [11] respectively. But most of these gating styles have their own demerits which had to be overcome as clock gating is one of the best ways to minimize the clock power dissipation. The modification approach for these demerits was reported in [4] with LECTOR–based clock gating (LB-CG), which is inspired by AND–based clock gating style and XOR–based comparator comprising of LECTOR–AND, LECTOR–INVERTER and static–CMOS XOR. One of the major demerits of the gating styles observed in [9–11] is the inability to stop the contention current during the transition of data line. This was successfully overcome in the LECTOR based gating approach. As a result, even if the data switches frequently or during its transition, the LECTOR will help stopping the contention current flowing in the clock gating logic. So, we have chosen LB – CG to drive multi–stage architecture application.

## 3. Circuit incorporating Multi–Stage Operation

The multi–stage circuit representation of register is a highlight in this paper. As flip flop is one bit storage element, therefore, multi–stage circuit design requires flip–flops. Basically, we have implemented a 2-bit register, which can be considered as two stage sequential circuit as shown in figure 2. In [4], a LB-CG based conventional D–Flip–Flop was already demonstrated and analyzed. The register is made of this LB-CG based D flip–flop. The data bit D1 and D2 of Register 1 are sent to Register 2 comprising of Q1 and Q2. In this course, the system clock is inserted as the global clock pulse common to both the flip flops. But, the clock gating is conducted individually. Therefore, ckg and ckg_1 in figure 2 will have different operating frequency. Instead, we could have used common gated clock in the architecture, but that may have introduced logical error. The reason is D1 and D2 changes continually, which will allow gated clock to be changed accordingly, as it is produced by the XOR comparison output of the MASTER Latch part. The architecture is simulated using 90nm PTM technology. Figure 3 displays the transient response of the architecture.

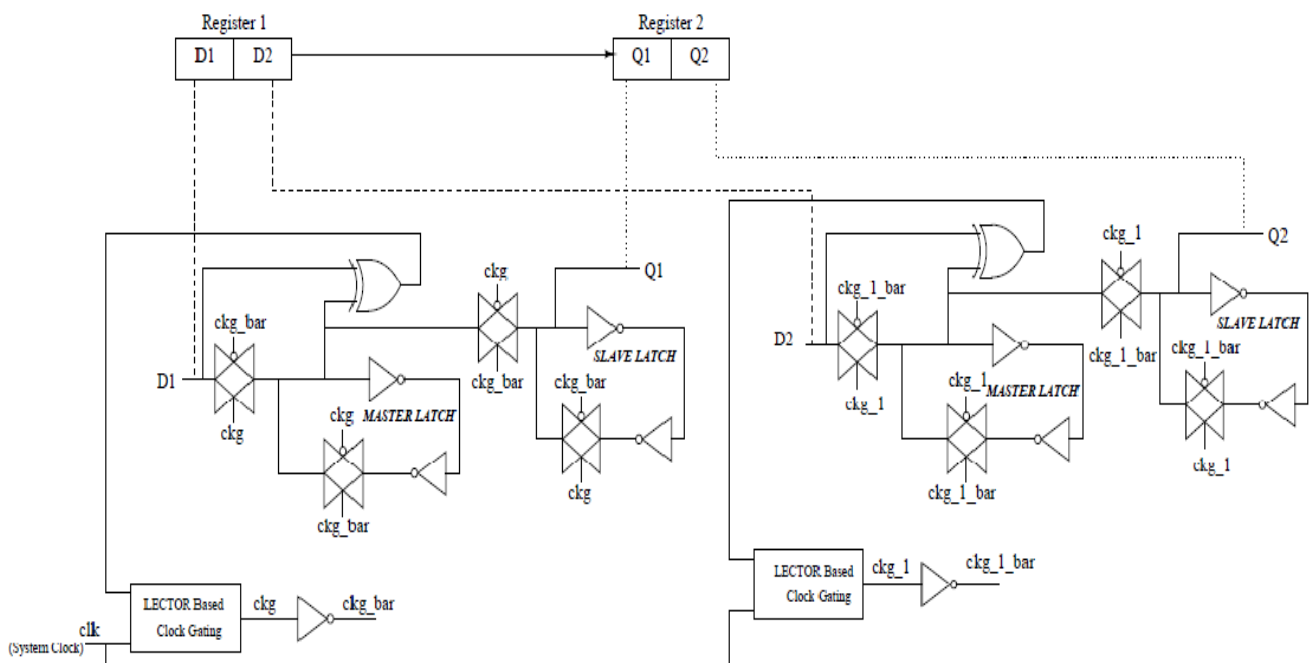

**Figure 2. Two-Stage Sequential Circuit: A Register**

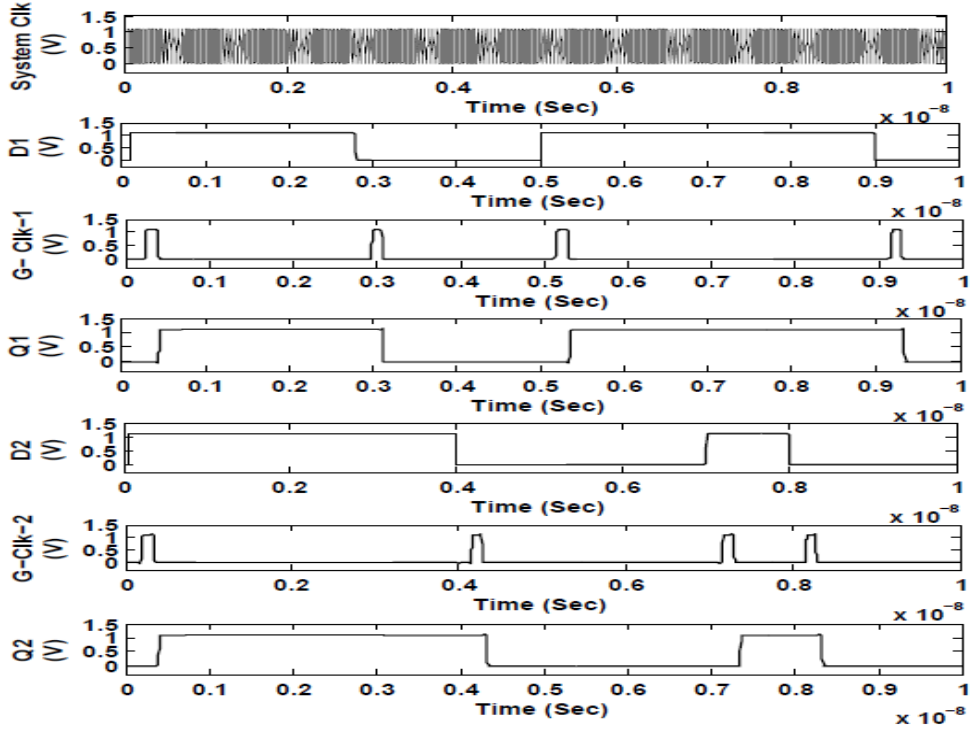

**Figure 3. Transient – Response of Multi-Stage Flip Flop architecture.**

## 4. Simulation Result & Analysis

In this section, we have discussed the simulation results obtained on simulating multi–stage architecture in SPICE using LB-CG technique. The analyses are sub–divided into three parts as (a) timing analysis, (b) power analysis and (c) process variation.

(a) For the timing analysis, the flip flop timing parameters has been estimated using the test circuit shown in figure 4. The propagation delay of the clock gated D Flip Flop in the multi–stage architecture, measured as the time interval between ckg and Q_flip_flop, during rising and falling transitions is

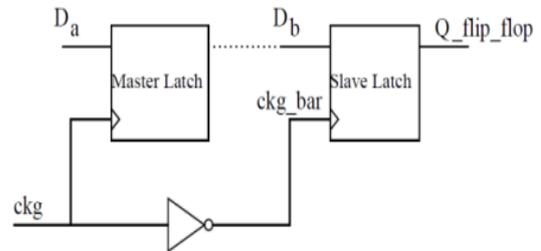

**Figure 4. Test Circuit for timing analysis.**

**Table 1: Timing Analysis of Single–stage from Multi–stage Gated Architecture**

| PARAMETERS | THIS WORK | | DOUBLE – GATED FLIP FLOP [10] | NC²MOS – GATED FLIP FLOP [10] |
|---|---|---|---|---|
| | WITH LB-CG | WITH NO CLOCK GATING | | |
| Setup (ns) | 0.047 | 0.045 | 1.40 | 1.07 |
| Hold (ns) | -0.16 | -0.028 | -1.04 | -1.01 |
| Delay (ns) | 1.46 | 0.154 | 1.35 | 0.98 |
| Latency (ns) | 1.507 | 0.199 | 2.75 | 2.05 |

0.213ns and 2.708ns respectively.

The measure of setup and hold time is based on the time interval between the positive edges of ckg_bar and the $D_b$, where ckg_bar is complimented counterpart of ckg. Setup time is defined as the time duration for the data $D_b$ to get stable before the ckg_bar arrives at the slave latch. Similarly, hold time is defined as the time duration for the data $D_b$ to remain unchanged till the ckg_bar switches. Latency is calculated as the summation of Setup and Delay. In Table 1, we have presented the timing analysis of gated multi–stage design in comparison with its non–gated correspondent.

(b) The power analysis shown in Table 2, depicts that the average power dissipation in gated clock approach of the architecture is qualitatively low in comparison to non–gated approach, even though it has been introduced extra circuit overhead. The average power dissipated by the architecture is 56.43µW at an operating frequency of 18GHz. The same logic is also simulated without gating at the same frequency and it offers an average power of 61.193µW.

**Table 2: Power Analysis of Multi–Stage Gated Architecture**

| PARAMETERS | WITH LB-CG | WITHOUT CLOCK GATING |
|---|---|---|
| Process Technology (nm) | 90 | 90 |
| Average Power (µW) | 56.43 | 61.19 |
| Transistor Count | 100 | 42 |
| Clock Frequency (GHz) | 18 | 18 |

(c) The multi–stage gated architecture has been simulated at a temperature of $30^0C$, with stable functionality and gives stable transient response to a process variation of 2%.

## 5. Conclusion

Power optimization of integrated circuits is largely possible with clock gating implementation in multi-stage sequential logics. So, employment of a proper clock gating technique is a need of an hour. As LECTOR based clock gating architecture has shown potential in literature, we assumed it to be a better option for sequential logics like counter, registers or any benchmark circuits. An implementation of 2-stage clock gated architecture has been shown in this paper. This implementation has offered a savings of 7.69% in average power with respect to its non–gated architecture with a penalty in transistor count and compromising timing performance. But, the architecture provides better timing performance in compared to the existing gating techniques.

## Acknowledgement

We are thankful to Ministry of Electronics & Information Technology, Government of India, for providing the financial grant under SMDP-C2SD project and Visvesaraya PhD Scheme.